\documentclass[]{aastex631}

\usepackage{amsmath} 
\newcommand\revision[1]{#1}

\begin{document}

\title{Col-OSSOS: Investigating the Origins of Different Surfaces in the Primordial Kuiper Belt}

\correspondingauthor{Laura E. Buchanan}
\email{laurabuchanan@uvic.ca}

\author[0000-0002-8032-4528]{Laura E. Buchanan}
\affiliation{Astrophysics Research Centre, School of Mathematics and Physics, Queen’s University Belfast, Belfast BT7 1NN, UK}
\affiliation{Department of Physics and Astronomy, University of Victoria, Elliott Building, 3800 Finnerty Road, Victoria, BC V8P 5C2, Canada}

\author[0000-0003-4365-1455]{Megan E. Schwamb}
\affiliation{Astrophysics Research Centre, School of Mathematics and Physics, Queen’s University Belfast, Belfast BT7 1NN, UK}

\author[0000-0001-6680-6558]{Wesley C. Fraser}
\affiliation{NRC-Herzberg Astronomy and Astrophysics, National Research Council of Canada, 5071 West Saanich Road, Victoria, BC V9E 2E7, Canada}

\author[0000-0003-3257-4490]{Michele T. Bannister}
\affiliation{School of Physical and Chemical Sciences—Te Kura Matū, University of Canterbury, Private Bag 4800, Christchurch 8140, New Zealand}

\author[0000-0001-7032-5255]{J. J. Kavelaars}
\affiliation{NRC-Herzberg Astronomy and Astrophysics, National Research Council of Canada, 5071 West Saanich Road, Victoria, BC V9E 2E7, Canada}
\affiliation{Department of Physics and Astronomy, University of Victoria, Elliott Building, 3800 Finnerty Road, Victoria, BC V8P 5C2, Canada}

\author[0000-0001-8617-2425]{Michaël Marsset}
\affiliation{European Southern Observatory, Alonso de Córdova 3107, Santiago, Chile}
\affiliation{Department of Earth, Atmospheric and Planetary Sciences, MIT, 77 Massachusetts Avenue, Cambridge, MA 02139, USA}

\author[0000-0003-4797-5262]{Rosemary~E. Pike}
\affiliation{Center for Astrophysics $|$ Harvard \& Smithsonian, 60 Garden Street, Cambridge, MA 02138, USA}

\author[0000-0002-4547-4301]{David Nesvorný}
\affiliation{Department of Space Studies, Southwest Research Institute, 1050 Walnut St., Suite 300, Boulder, CO, 80302, United States}

\author[0000-0001-5368-386X]{Samantha M. Lawler}
\affiliation{Campion College and the Department of Physics, University of Regina, Regina, SK, Canada S4S 0A2}

\author[0000-0001-8821-5927]{Susan D. Benecchi}
\affiliation{Planetary Science Institute, 1700 East Fort Lowell, Suite 106, Tucson, AZ 85719, USA}

\author[0000-0002-6830-476X]{Nuno Peixinho}
\affiliation{Instituto de Astrof\'{\i}sica e Ci\^{e}ncias do Espa\c{c}o, Instituto de Investigação Interdisciplinar, Universidade de Coimbra, 3040-004 Coimbra, Portugal}

\author[0000-0001-6541-8887]{Nicole J. Tan}
\affiliation{School of Physical and Chemical Sciences—Te Kura Matū, University of Canterbury, Private Bag 4800, Christchurch 8140, New Zealand}

\author[0000-0001-8736-236X]{Kathryn Volk} 
\affiliation{Planetary Science Institute, 1700 East Fort Lowell, Suite 106, Tucson, AZ 85719, USA}

\author[0000-0003-4143-8589]{Mike Alexandersen}
\affiliation{Center for Astrophysics $|$ Harvard \& Smithsonian, 60 Garden Street, Cambridge, MA 02138, USA}

\author[0000-0003-0407-2266]{Jean-Marc Petit}
\affiliation{Institut UTINAM UMR6213, CNRS, Univ. Bourgogne Franche-Comté, OSU Theta F-25000 Besançon, France}

\begin{abstract}

    The Colours of the Outer Solar System Origins Survey (Col-OSSOS) measured the optical/NIR colours of a brightness-complete sample of Trans-Neptunian Objects (TNOs). Like previous surveys, this one found a bimodal colour distribution in TNOs, categorised as red and very red. Additionally, this survey proposed an alternative surface classification scheme: FaintIR and BrightIR. Cold classical TNOs mostly have very red or FaintIR surfaces, while dynamically excited TNOs show a mixture of surfaces. This likely indicates that formation locations and proximity to the Sun influenced surface characteristics and color changes. Our study combines the data from Col-OSSOS with two dynamical models describing the formation of the Kuiper belt during Neptune's migration. We investigate the proposed surface-colour changing line and explore the distribution of different surfaces within the primordial disk. By comparing radial colour transitions across various scenarios, we explore the origins of surface characteristics and their implications within the context of BrightIR and FaintIR classifications. Moreover, we extend our analysis to examine the distribution of these surface classes within the present-day Kuiper Belt, providing insights into the configuration of the early solar system's planetesimal disk prior to giant planet migration. We find that the most likely primordial disk compositions are inner neutral / outer red (with transition $30.0^{+1.1}_{-1.2}$ au), or inner BrightIR / outer FaintIR (with transition $31.5^{+1.1}_{-1.2}$ au).

\end{abstract}


\section{Introduction} \label{sec:intro}

    In the outer solar system beyond Neptune lies the Kuiper belt, made up of a region of small icy bodies known as Trans-Neptunian Objects (TNOs). As these TNOs have remained largely undisturbed since the end of the giant planet migration period, studying them can provide valuable insights into the conditions of the early solar system in which they formed. TNOs can be split into two main subpopulations, the dynamically cold TNOs and the dynamically hot (or excited) TNOs. The dynamically excited TNOs formed further sunward than they are today, and then were emplaced onto their current orbits by a period of giant planet migration \citep[e.g.,][]{2005Natur.435..459T,2005Natur.435..462M,2006AJ....131.1142S,2007AJ....133.1962N,2008Icar..196..258L}.  The dynamically excited TNOs (with $r$ mag $>\sim22$) exhibit a bimodal distribution of surface colours, often referred to as red and neutral coloured TNOs \citep{2012ApJ...749...33F, 2012A&A...546A..86P, 2015A&A...577A..35P, 2015ApJ...804...31F, schwamb_col-ossos:_2019,2023PSJ.....4...80F}. 

\begin{figure}[ht]
        \centering
        \includegraphics[width=\textwidth]{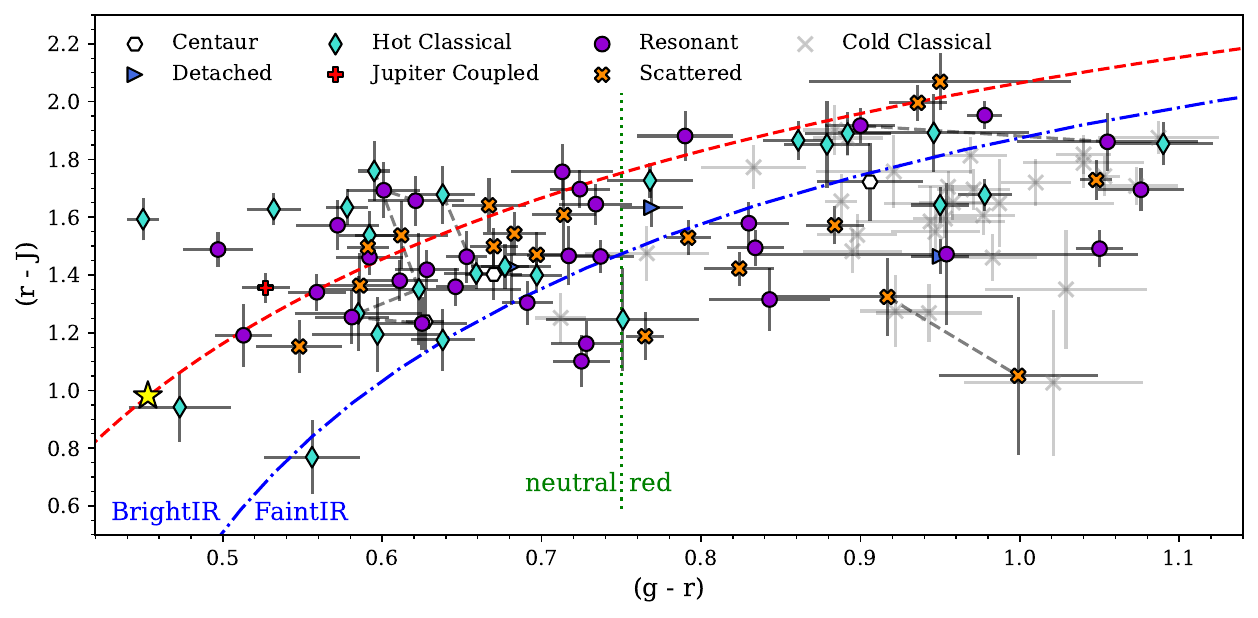}
        \caption{Optical and NIR colours of the full Col-OSSOS dataset with 102 TNOs. The shapes and colours of each point show the dynamical classifications, as described in the legend. The solar colours, with $g-r=0.45$ and $r-J=0.97$, are shown by the yellow star. The red dashed line shows the solar reddening line, while the blue dot-dashed line shows the $PC^2$ $= -$0.13 split between BrightIR and FaintIR surfaces. The vertical green dotted line shows the well known bifurcation in the optical colour distribution of the dynamically excited populations at $(g-r) = 0.75$. For TNOs with multiple surface colour observations in the survey, the colour points are joined with dashed lines. Some of these TNOs only had repeat measurements in the optical $(g-r)$, in which case the same $(r-J)$ colour is used for both and the missing NIR colour indicated by a lack of error bars.}
        \label{fig:colossosallcolours}
    \end{figure}

    The Colours of the Outer Solar System Origins Survey \citep[Col-OSSOS,][]{schwamb_col-ossos:_2019,2023PSJ.....4...80F} surveyed the optical and near-infrared (NIR) colours of a brightness-complete sample of 102 TNOs.  \citet{schwamb_col-ossos:_2019} proposed that the early observations of Col-OSSOS showed a bifurcation in the surface colours, split into neutral surfaces at $(g-r) < 0.75$ and red surfaces at $(g-r) \geq 0.75$. Additionally, \citet{2023PSJ.....4...80F} demonstrated it is more likely that the surface types exhibited by TNOs is not truly reflected in the bimodality of the optical colour distribution, but instead follows a bifurcated continuum in optical-NIR colours. TNOs along the solar reddening line have a 1:1 ratio between their optical and NIR spectral slopes, essentially meaning a single linear reflectance from the optical to NIR wavelengths. Therefore the TNOs above the reddening line have convex reflectance spectra, while those below the line have concave. \citet{2023PSJ.....4...80F} showed that in this $PC^1$ (distance along the solar reddening line), $PC^2$ (perpendicular distance from the solar reddening line) projection space the TNO surfaces can be split into two classifications (FaintIR and BrightIR) with the split between located at a $PC^2$ value of $-0.13$. This alternative surface taxonomy examines the surfaces of TNOs at optical and NIR wavelengths, in contrast to the historical red/neutral classification based solely on optical wavelengths. Therefore in the context of these IR classes, the BrightIR and FaintIR groups are each made up of a mixture of red and neutral surfaces. Both of these surface classifications are shown in Figure \ref{fig:colossosallcolours}.

    It is hypothesised that the colour distribution we see today originated from the primordial disk that gave rise to the Kuiper belt, potentially through a surface colour changing ice line \citep{2012ApJ...749...33F,2013Icar..222..307D}. These TNOs in the primordial disk were later emplaced into the outer Solar System \citep[where the cold classical TNOs formed,][]{2010ApJ...722L.204P} by the migration of the giant planets. However, while their large distance from the Sun limits certain forms of surface alteration, their surfaces have likely undergone processing through various mechanisms, including space weathering from irradiation \citep[e.g.,][]{2006ApJ...644..646B, 2008ssbn.book..483D, 2023Icar..39415396Q}, which contributes to the observed red spectral component. For example, \citet{2024NatAs.tmp..305D} suggest that the loss of C-bearing ices on TNOs that formed closer to the sun would have prevented the later formation of complex hydro-carbons via irradiation, and therefore more neutral coloured surfaces for those TNOs that lost those ices. Additionally, findings from New Horizons suggest that collisions may also play a role in modifying surface properties \citep[e.g.,][]{2022JGRE..12707068K}. Given these factors, while surface colours can provide clues about a TNO's original formation environment, they also reflect the cumulative effects of post-formation processing.
    
    Previously, the colours of the Kuiper belt we see today have been used to investigate the position of this colour / surface transition. \citet{2020AJ....160...46N} used resonant TNO data to measure possible transition locations under an assumed dynamical history, finding that with an inner neutral / outer red disk that spanned from $\sim$24 au to $\sim$50 au the transition position could lie between 30 au and 40 au. Following this \citet{2022PSJ.....3....9B} combined the initial data release of the Col-OSSOS, together with the dynamical model of \citet{nesvorny_neptunes_2016} to study primoridal disk with both an inner red / outer neutral and inner neutral / outer red layouts. By using the number of different surfaces in different regions of the sky observed by Col-OSSOS they found transition positions of $28^{+2}_{-3}$ au for the former and $27^{+3}_{-3}$ au for the latter best matched the numbers of each surface colour observed by Col-OSSOS at that time. 

    In this work we are developing further the work of \citet{2022PSJ.....3....9B}, combining together the full Col-OSSOS release \citep{2023PSJ.....4...80F} with two dynamical models of the formation of the Kuiper belt through Neptune's migration; \citet{nesvorny_neptunes_2016} and \citet{2020AJ....160...46N}. While future surveys such as the Legacy Survey of Space and Time \citep[LSST,][]{2009arXiv0912.0201L,2019ApJ...873..111I} will expand the sample of observed TNOs, they will not necessarily provide the photometric precision or wavelength coverage required to fully address this problem. Conducting this analysis now gives us a clear starting point for comparing models, helping to refine our understanding of which disk structures and migration histories are most plausible. This work provides important constraints on which models remain viable, ensuring that when new data becomes available, it can be interpreted in a clear and consistent way, rather than in isolation. The two migration models we use here have different primordial disks, allowing us to probe different formation scenarios. Additionally, we adopt an alternative statistical method to \citet{2022PSJ.....3....9B}, using binomial statistics rather than matching the exact observed number of each surface. In Section \ref{sec:colourtransitions} we outline our simulations exploring colours in the primordial Kuiper Belt along with their results and in Section \ref{sec4:L7sims} we use the CFEPS L7 model to make an estimate of the intrinsic colours within the Kuiper Belt. Finally, in Section \ref{sec4:conclusions} we summarise the work and make concluding remarks.

\section{Simulating Colour Transitions in the Primordial Kuiper Belt}\label{sec:colourtransitions}

    This work employs two dynamical models of the Kuiper belt's formation throughout the migration of Neptune \citep[][outlined in Sections \ref{ssec4:models_nes16} and \ref{ssec4:models_nes20}]{nesvorny_neptunes_2016,2020AJ....160...46N} in order to examine the radial position of a surface transition line (or `simulated ice line') in the primordial Kuiper belt, and how it relates to the colours and surface types observed in the Kuiper belt today. These are two separate viable dynamical histories, with one lacking excited progenitor objects outside of 30 au and so requires the surface transition to occur inside 30 au. We step the colour/surface transition through the range of transition positions for each migration model, so the post-Neptune migration simulated TNOs have surfaces assigned based on their starting positions. To ensure consistent and reliable comparisons between the simulation results and the dynamically excited populations observed in Col-OSSOS, we apply a set of dynamical constraints to both the dynamical models and the observed TNOs. By utilising the OSSOS survey simulator \citep{lawler_ossos:_2018}, we can directly compare the simulated TNOs with the observations from Col-OSSOS, allowing us to determine the most plausible compositions of the primordial disk that could give rise to the observed Kuiper belt. We use binomial probabilities to assess the likelihood of attaining the observed outcomes using the simulated surface populations. Through adjusting the position of the surface transition within the pre-Neptune migration disk, we investigate the optimal location that best reproduces the colours observed by Col-OSSOS in the present-day Kuiper belt.

    Previous work by \citet{2022PSJ.....3....9B} explored the presence of an inner red / outer neutral disk as well as an inner neutral / outer red disk. Within the Kuiper belt `blue binaries' (neutral-coloured wide-set TNO binaries) were believed to have formed at ~$\sim$38 au \citep{fraser_all_2017}, and could not have migrated significant distances of more than ~$\sim$5 au without losing their binarity. In the case of \citet{2022PSJ.....3....9B} the primordial disk spanned 24 au $-$ 30 au and so neutral-coloured TNOs would have had to form further out in the disk than those with redder surfaces for the blue-binaries to exist. This provided observational basis to test the inner red / outer neutral disk layout. Some observation evidence for the inner neutral / outer red initial disk was the fact that the majority of known Neptune Trojan colours at the time were neutral \citep[16 out of 17 were neutral][]{2013AJ....145...96P,2018AJ....155...56J}, consistent with the notion that neutral objects formed further inward. Since that work was conducted, there has been a significant increase in the number of known Neptune Trojan colours. Specifically, \citet{2023MNRAS.521L..29B} measured the colours of an additional 15 Neptune Trojans and discovered four more red surfaces, implying that a more sunward transition from red to neutral is needed to account for the distribution of surface colours in the Neptune Trojans. \citet{2023MNRAS.521L..29B} suggest that a transition inward of $30 - 35$ au is necessary to account for this distribution, although they only discussed the case of an inner neutral / outer red disk. 
    
    The two dynamical models we use in this work are \citet{nesvorny_neptunes_2016} and \citet{2020AJ....160...46N}. Although they both have similar conditions for the migration of Neptune, the planetesimal disks from which their simulated Kuiper belts originate differ. For the \citet{nesvorny_neptunes_2016} migration model the initial disk spans between $\sim$24 au and $\sim$ 30 au. For the \citet{2020AJ....160...46N} migration model there is a similar condensed disk between $\sim$24 au and $\sim$ 30 au, with the addition of an extended lower mass disk from $\sim$30 au to $\sim$ 50 au. 
    
    \subsection{Condensed Disk Migration Model} \label{ssec4:models_nes16}
        
\begin{figure}[ht]
            \centering
            \includegraphics[width=\textwidth]{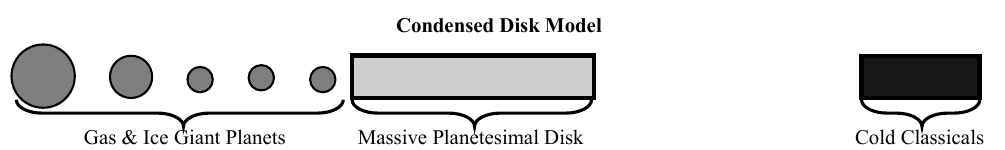}
            \caption{A schematic of the disk in the model of \citet{nesvorny_neptunes_2016}, where the massive planetesimal disk spans from $\sim$24 au to $\sim$30 au. Although the cold classical TNOs are not included in the simulation, a representation of their outer solar system position (beyond $\sim$50 au) position is shown. In this dynamical model there is no contribution material between the edge of the massive planetesimal disk at $\sim$30 au and the cold classical TNOs.}
            \label{fig:disk2016}
        \end{figure}

        The first migration model we use is that of \citet{nesvorny_neptunes_2016} (here referred to as the condensed disk model), which involved an N-body simulation depicting the formation of the Kuiper belt. This simulation accurately tracked the trajectories of the simulated TNOs from their initial positions to their ultimate orbits. The dynamical model consisted of five giant planets and a slow migration of Neptune with an e-folding timescale of 10 Myrs, where the migration is grainy due to scattering with massive planetesimals within the disk throughout its migration. Additionally, as Neptune migrates outwards from $<25$ au $-$ $\sim30$ au it has a `jump' of $\sim 0.5$ au at 28 au. The initial disk in this model spanned approximately 24 au to 30 au with one million test particles making up an estimated disk mass of 20$M_{Earth}$, and in Figure \ref{fig:disk2016} we show a schematic of the pre-migration disk layout.  This particular model of Kuiper belt formation demonstrates a reasonable level of accuracy in replicating the characteristics of the current Kuiper belt. Notably, it effectively addresses the issue of overpopulated resonances that had plagued earlier dynamical models \citep[e.g.,][]{2005AJ....130.2392H, 2008Icar..196..258L, 2008ssbn.book..275M}.
            
    \subsection{Extended Disk Migration Model} \label{ssec4:models_nes20}
        
\begin{figure}[ht]
            \centering
            \includegraphics[width=\textwidth]{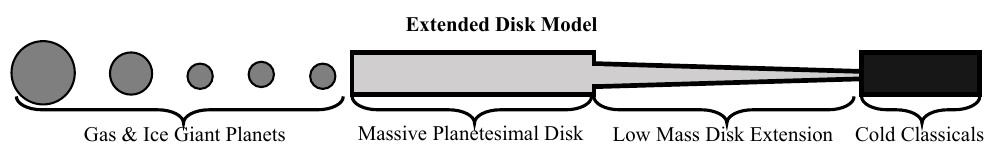}
            \caption{A schematic of the disk in the model of \citet{2020AJ....160...46N}, where a massive planetesimal disk spans from $\sim$24 au to $\sim$30 au and beyond this is a lower mass disk extension that spans to near the cold classical region. Although the cold classical TNOs are not included in the simulation, a representation of their outer solar system position (beyond $\sim$50 au) position is shown.}
            \label{fig:disk2020}
        \end{figure}

        The second model that we use in this work is by \citet{2020AJ....160...46N}, and here referred to as the extended disk model. It consists of a similar dense disk (along with five giant planets) to that of the condensed disk model above between $\sim$24 au and $\sim$30 au, and then an extension beyond this with a lower density of simulated TNOs as demonstrated in Figure \ref{fig:disk2020}. This model allows investigations with TNOs that formed at further out distances, while keeping the density beyond 30 au low enough that Neptune's outward migration stops at its current orbit. 
        
        In this simulation Neptune starts at 24 au and migrates outwards for 0.5 Gyrs. The first stage of this has an e-folding exponential migration timescale ($\tau$) of 10 Myrs, then has a `jump' in its migration of 0.4 au before continuing outwards with  $\tau$ of 30 Myrs to its current orbit at 30.11 au. The primordial disk is made up of $2 \times 10^6$ simulated TNOs, with a total disk mass of 20 $M_{Earth}$. Within this disk are 2000 Pluto-mass particles, throughout its migration Neptune interacts with these causing a `graininess' and aiding in reducing the number of TNOs that are left in the MMRs. There are a number of disk profiles tested by \citet{2020AJ....160...46N} (exponential, truncated, and a hybrid of the two), and in this case we are using the truncated disk profile as it best matches the orbital structure within the Kuiper belt for this migration timescale. The profile is described by \(\Sigma \propto r^{-\gamma}\). Here $r$ is the radial distance through the disk and $\gamma = 2$. Beyond 30 au there is a step down in surface density parameterised by the contrast $c$ (where \(\Sigma \propto c^{-1}r^{-\gamma}\)), equal to 1000 for this work \citep[from][]{2020AJ....160...46N}. The significant drop in disk density at 30 au is essential to ensure that Neptune stops its migration on an orbit of $\sim$30 au, while the low mass extension means that there is material in the initial disk that extends to near the cold classical formation region.
        
    \subsection{Dynamical Cuts to the Model} \label{ssec4:dyn_cuts}
    
        For this work we are only investigating TNOs that were emplaced onto their current orbits by the migration of Neptune. We also want to ensure that we make accurate comparisons between the model and observations, therefore we employ consistent dynamical criteria for both, guaranteeing that any synthetic TNOs within our generated populations would have also been detectable by Col-OSSOS. Therefore, we outline below how and why we remove the cold classical TNOs, those on MMRs and the Centaur populations, along with any TNOs further out that will have significantly evolved orbits beyond the end of Neptune's migration. Although the majority of the dynamical cuts to the condensed and extended disk models are identical, we treat the MMRs differently. This is discussed in Section \ref{nes16_MMRs}.
        
        A more modern definition for cold classical orbits uses a `free inclination' cut of $4^\circ$, where the free inclination is the orbit's inclination relative to a locally dynamically meaningful reference plane \citep{2022ApJS..259...54H}. Although these free inclinations are available for the observed TNOs, they cannot be accurately computed for the simulated TNOs. This would require significant further integrations of the orbits in order to calculate their distances from the $\nu_{18}$ secular resonance \citep{2022ApJS..259...54H}. Therefore, we are using the ecliptic inclinations which have historically been used to define the cold classicals. Using the classical belt orbits defined by \citet{gladman_nomenclature_2008}, we categorise classical TNOs as those with semimajor axis range 37.37 au $< a <$ 47.7 au and eccentricities below 0.24. We therefore remove any synthetic TNOs within these limits, with inclinations less than 5$^\circ$. 
        
        Accurately tracing the dynamical history of observed TNOs on the major MMRs is not reliably achievable. Their orbital evolution in the 4 Gy since the end of Neptune's migration takes away the ability to accurately trace their origin positions. Additionally, cold classical TNOs can be captured and diffused into the MMRs, and therefore contaminate the surface populations \citep{2019AJ....158...53T}. Accordingly we remove the major MMRs from both the simulations and the observations. The different treatments for MMR removal are covered in Section \ref{nes16_MMRs}.
        
        Centaurs have shorter dynamical lifetimes, and so have likely significantly evolved in their orbits since the end of Neptune's migration \citep{2013Icar..224...66V,2020CeMDA.132...36D}. Additionally, Centaurs pass closer to the sun on their orbits, and so their surfaces can undergo processing and therefore may no longer be primordial \citep{2020tnss.book..307P}. We therefore define a Centaur orbit classification, based on that of \citet{gladman_nomenclature_2008} and remove these orbits from both the simulations and the observations. For this adapted definition we use a semimajor axis less than that of Neptune, aphelion distance $>$11 au. Finally we include the condition that the simulated TNOs must have semimajor axis less than 250 au and a perihelion distance $<$45 au. TNOs with orbits beyond these limits have likely significantly evolved since the end of Neptune's migration \citep{2015MNRAS.446.3788B}.

        \subsubsection{Handling of the MMRs}\label{nes16_MMRs}
        
            The final orbits of the simulated TNOs in the condensed disk model were forward integrated. Therefore the TNOs on the major MMRs could be identified. We remove the simulated TNOs on the 3:2, 5:2, 4:3, 5:3, 7:4, and 2:1 MMRs identified by \citet{nesvorny_neptunes_2016}. The final orbits of the extended disk model have not been forward integrated and so the MMRs have not been dynamically identified. We instead use a semimajor axis cut for the three strongest MMRs. Although there are additional MMRs in the classical region of the Kuiper belt \citep[e.g. 5:2 and 7:4 MMRs which could hold cold classical TNOs up to inclinations of $10^\circ$,][]{2024_Thirouin}, the 2:1, 3:2 and 4:3 MMRs are the most populous in this region. In fact, within the Col-OSSOS observations no other MMRs have members in the classical belt with orbital inclinations $5^\circ<i<10^\circ$, where cold classical contamination from MMRs is most likely to occur. Additionally, these MMRs are reasonably isolated, allowing us to remove the simulated TNOs that would potentially be on MMR orbits without significantly losing additional non-resonant TNOs. We reject from the sample any objects within ~$\pm$0.5 au of the centres of each of the main MMRs at 39.45, 47.8, and 36.5 au.

    \subsection{Col-OSSOS Comparison Sample}\label{ssec4:col_n}
    
\begin{figure}[ht]
            \centering
            \includegraphics[width=\textwidth]{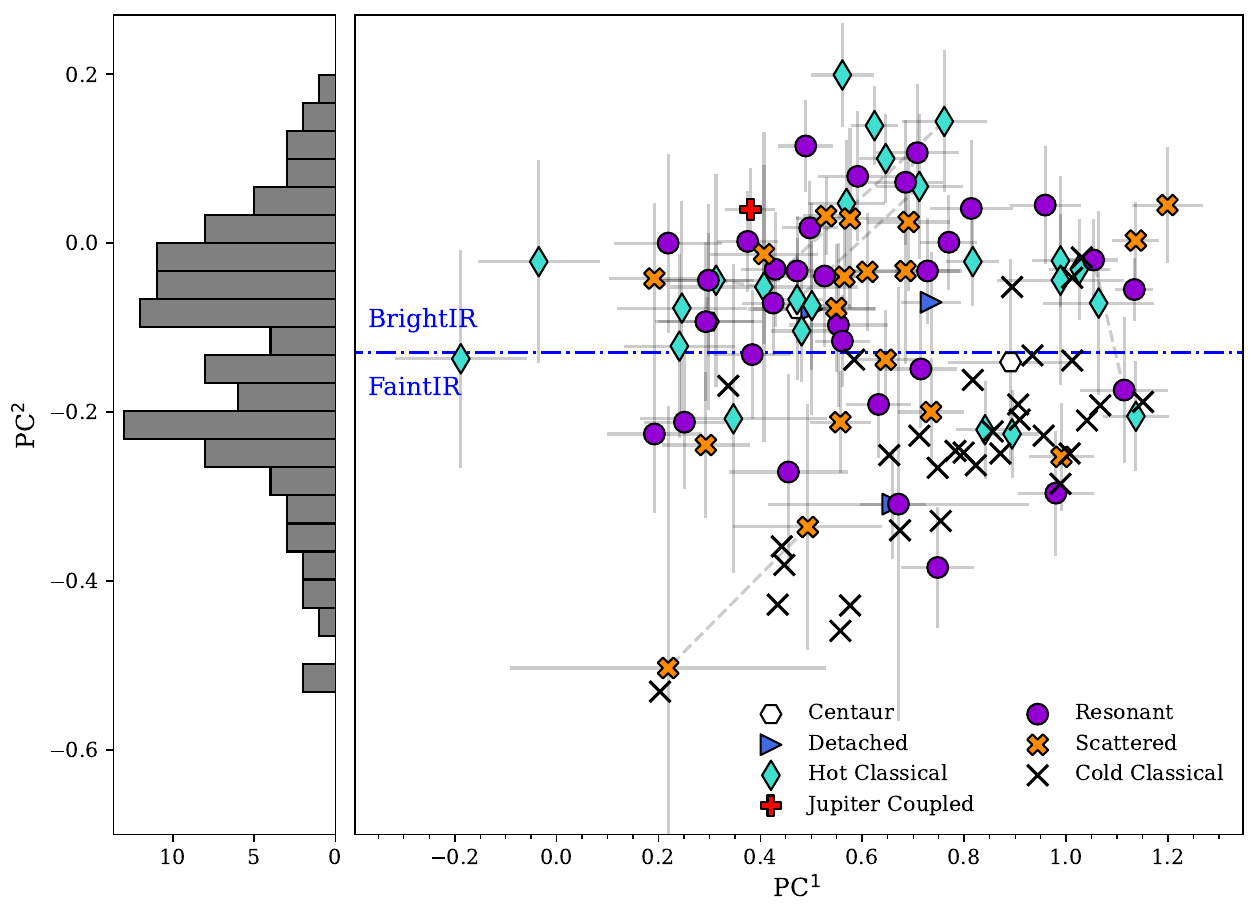}
            \caption{The projection values $PC^1$ and $PC^2$ of the $(g - r)$ and $(r - J)$ colours of Col-OSSOS in the right-hand panel. In this projection, the reddening curve would be a horizontal line with $PC^2 = 0$. The blue dot-dashed line is at a value of $PC^2$ $= -$0.13, the value the split between the FaintIR and BrightIR classes. Similarly to Figure \ref{fig:colossosallcolours}, TNOs with multiple measurements have dashed lines between their points. The left-hand panel shows a histogram of the distribution in PC$^2$, illustrating the bimodal distribution from which the BrightIR and FaintIR classes are defined.}
            \label{fig:colossospcs}
        \end{figure}

        Col-OSSOS observations were acquired between 2014 and 2022, measuring near simultaneous $g-$, $r-$, and $J-$band photometry of a subset of TNOs from the the Outer Solar System Origins Survey \citep[OSSOS,][]{bannister_outer_2016,bannister_ossos._2018}. Col-OSSOS selected TNOs with $r-$band magnitude $<23.6$ within the E, H, L, O, S and T OSSOS observing blocks. \revision{Each observing block corresponds to a distinct and non-overlapping region of the sky. The measurements obtained in one block are statistically independent of those in any other, as each region samples a unique portion of the survey footprint with its own detection completeness and colour distribution.} The initial data releases are published in \citet{schwamb_col-ossos:_2019} and \citet{2022PSJ.....3....9B}, while the full survey is published in \citet{2023PSJ.....4...80F}. The full Col-OSSOS sample is described in \citet{2023PSJ.....4...80F} and consists of a brightness-complete sample of 102 TNOs with optical and NIR colours measured to an exceptionally high precision. In Figure \ref{fig:colossosallcolours} we show the $(g - r)$ and $(r - J)$ colours while in Figure \ref{fig:colossospcs} we show the $PC^1$ and $PC^2$ of the $(g - r)$ and $(r - J)$ colours.
        
        Here we explore all six observing blocks of Col-OSSOS, rather than just the three used in \citet{2022PSJ.....3....9B} as all six blocks have since been completed. In this section we will summarise the subset of TNOs that we use for comparison with the dynamical models. We refer to these subsets of the observations as the condensed disk comparison sample and the extended disk comparison sample. We use the OSSOS Survey Simulator \citep{lawler_ossos:_2018} to bias the simulated TNOs to what OSSOS could have detected. As Col-OSSOS observed OSSOS targets with apparent $r-$band magnitudes brighter than 23.6 within the E,L,H,O,S and T observing blocks we limit the survey simulator outputs to these same conditions. Additionally, Col-OSSOS was selecting targets as OSSOS was still refining its characterization procedure, and so two objects ended up with $r-$band detection magnitudes fainter than 23.6. As we are applying a 23.6 magnitude cut to the simulations, these sources are removed from further consideration.

        Within Col-OSSOS 19 TNOs had multiple colour measurements. Only one of these TNOs has repeat measurements that fall on either side of the neutral/red class boundary, 2013 JE64. However, the initial observation of this TNO was possibly impacted by a bright background star and therefore unreliable, so we use the follow-up optical photometry to classify the TNO as neutral coloured. For the rest of the repeat observed TNOs we simply use their initial photometry, as they do not change colour/surface classes between observations. Finally, we have a single member of the Haumea collisional family within Col-OSSOS \citep{pike_col-ossos_2017}. These are the remaining fragments of a long ago collision with the dwarf planet Haumea \citep{2007AJ....133..284B,2008ApJ...684L.107S,2010A&A...511A..72S,2011ApJ...730..105T,2012A&A...544A.137C,2012ApJ...749...33F,2019AJ....157..230P}. They are distinguished based on their neutral surface colours, water ice rich surfaces and clustered orbital properties. As the surface of this TNO is the result of a collision and is not primordial, we remove 2013 UQ15 from the observed sample.

        In addition, we must make dynamical cuts to this comparison sample. This is to ensure that we are only comparing dynamically excited TNOs that were emplaced onto their current orbits by the migration of Neptune. By removing these TNOs via an orbital cut rather than simply using the OSSOS dynamical classifications it allows us to use the same limits with the dynamical models. We apply the exact same orbit cut as those to the dynamical model, outlined in Section \ref{ssec4:dyn_cuts} and with the MMR removal covered in Section \ref{cs16_MMRs} below.

        In Figure \ref{fig:nes16comp} we show the colours of both the condensed disk comparison sample and the extended disk comparison sample. In Table \ref{tab:dynmod_sample} we show the colours and orbits of the TNOs along with their corresponding comparison sample. Finally in Table \ref{tab:n_nums} we summarise the total numbers of each surface type per observing block in the comparison sample, containing a total of 37 TNOs for the condensed disk model and 40 TNOs in the extended disk model. The difference here is simply due to the slightly different MMR removal for each case.
        
        \startlongtable
        \begin{deluxetable}{cccccccccccc}
            \tablecaption{\label{tab:dynmod_sample} Orbital Parameters and Optical and NIR colors of the Col-OSSOS observations compared with the two dynamical models of Neptune's migration outlined in Sections \ref{ssec4:models_nes16} and \ref{ssec4:models_nes20}. The TNO classifications are abbreviated with cen = centaur, sca = scattering disk, det = detached, cla = classical belt, res= MMR with Neptune. N16 refers to the TNO being a part of the \citet{nesvorny_neptunes_2016} comparison sample while N20 refers to the TNO being part of the \citet{2020AJ....160...46N} comparison sample.}
            \tablehead{
                \colhead{MPC} & \colhead{OSSOS ID} & \colhead{Classification} & \colhead{Mean $m_r$} & \colhead{$H_r$} & \colhead{a (au)} & \colhead{e} & \colhead{i ($^{\circ}$)} & \colhead{$(g-r)$} & \colhead{$(r-J)$} & \colhead{Model Sample} }
            \startdata
                \hline
                2002 GG166 & o3e01 & sca & $21.50\pm0.09$ & 7.73 & 34.42 & 0.59 & 7.71 & $0.59\pm0.01$ & $1.50\pm0.05$ & N16,N20 \\
                2013 GY136 & o3e09 & res & $22.94\pm0.05$ & 7.32 & 55.54 & 0.41 & 10.88 & $0.51\pm0.02$ & $1.19\pm0.11$ & N20 \\
                2001 FO185 & o3e23PD & cla & $23.37\pm0.08$ & 7.09 & 46.45 & 0.12 & 10.64 & $0.86\pm0.02$ & $1.87\pm0.07$ & N16,N20 \\
                2013 GO137 & o3e29 & cla & $23.46\pm0.08$ & 7.09 & 41.42 & 0.09 & 29.25 & $0.77\pm0.03$ & $1.73\pm0.06$ & N16,N20 \\
                2013 GP136 & o3e39 & det & $23.07\pm0.07$ & 6.42 & 150.24 & 0.73 & 33.54 & $0.77\pm0.02$ & $1.63\pm0.07$ & N16,N20 \\
                2013 GG138 & o3e44 & cla & $23.26\pm0.09$ & 6.34 & 47.46 & 0.03 & 24.61 & $1.09\pm0.03$ & $1.85\pm0.07$ & N16 \\
                2013 HR156 & o3e49 & res & $23.54\pm0.09$ & 7.72 & 45.72 & 0.19 & 20.41 & $0.59\pm0.03$ & $1.36\pm0.11$ & N16,N20 \\
                2013 GM137 & o3e51 & cla & $23.32\pm0.23$ & 6.90 & 44.10 & 0.08 & 22.46 & $0.60\pm0.04$ & $1.19\pm0.13$ & N16,N20 \\
                2013 GY137 & o3e53 & cla & $23.50\pm0.23$ & 7.29 & 44.89 & 0.10 & 5.31 & $0.88\pm0.03$ & $1.85\pm0.15$ & N16,N20 \\
                2001 QF331 & o3l06PD & res & $22.71\pm0.07$ & 7.56 & 42.25 & 0.25 & 2.67 & $0.83\pm0.03$ & $1.58\pm0.07$ & N20 \\
                2013 SZ99 & o3l15 & cla & $23.54\pm0.13$ & 7.65 & 38.28 & 0.02 & 19.84 & $0.59\pm0.02$ & $1.54\pm0.08$ & N16,N20 \\
                2010 RE188 & o3l18 & cla & $22.27\pm0.05$ & 6.19 & 46.01 & 0.15 & 6.75 & $0.68\pm0.01$ & $1.43\pm0.08$ & N16,N20 \\
                2013 UM17 & o3l29PD & cla & $23.56\pm0.09$ & 7.29 & 42.48 & 0.08 & 12.99 & $0.75\pm0.05$ & $1.25\pm0.18$ & N16,N20 \\
                2013 JK64 & o3o11 & res & $22.94\pm0.04$ & 7.69 & 55.25 & 0.41 & 11.08 & $0.90\pm0.02$ & $1.92\pm0.06$ & N20 \\
                2013 JO64 & o3o14 & sca & $23.54\pm0.08$ & 8.00 & 143.30 & 0.75 & 8.58 & $0.55\pm0.03$ & $1.15\pm0.09$ & N16,N20 \\
                2013 JR65 & o3o21 & cla & $23.51\pm0.12$ & 7.53 & 46.20 & 0.19 & 11.71 & $0.45\pm0.01$ & $1.59\pm0.07$ & N16,N20 \\
                2013 JN65 & o3o28 & cla & $23.42\pm0.21$ & 7.23 & 40.67 & 0.01 & 19.64 & $0.58\pm0.01$ & $1.63\pm0.06$ & N16,N20 \\
                2013 JL64 & o3o29 & det & $23.26\pm0.12$ & 7.03 & 56.77 & 0.37 & 27.67 & $0.68\pm0.03$ & $1.43\pm0.13$ & N16,N20 \\
                2013 JH64 & o3o34 & res & $22.70\pm0.04$ & 5.60 & 59.20 & 0.38 & 13.73 & $0.70\pm0.02$ & $1.47\pm0.08$ & N16,N20 \\
                2013 JX67 & o3o51 & cla & $22.75\pm0.04$ & 6.49 & 46.39 & 0.13 & 10.50 & $0.70\pm0.02$ & $1.40\pm0.06$ & N16,N20 \\
                2014 UQ229 & o4h03 & sca & $22.69\pm0.21$ & 9.55 & 49.90 & 0.78 & 5.68 & $0.94\pm0.02$ & $2.00\pm0.06$ & N16,N20 \\
                2014 US229 & o4h14 & res & $23.18\pm0.08$ & 7.95 & 55.26 & 0.40 & 3.90 & $0.63\pm0.02$ & $1.42\pm0.07$ & N20 \\
                2014 UK225 & o4h19 & cla & $23.23\pm0.06$ & 7.43 & 43.52 & 0.13 & 10.69 & $0.98\pm0.02$ & $1.68\pm0.06$ & N16,N20 \\
                2014 UL225 & o4h20 & cla & $23.03\pm0.07$ & 7.24 & 46.34 & 0.20 & 7.95 & $0.56\pm0.03$ & $0.77\pm0.13$ & N16,N20 \\
                2014 UH225 & o4h29 & cla & $23.31\pm0.06$ & 7.30 & 38.64 & 0.04 & 29.53 & $0.53\pm0.02$ & $1.63\pm0.06$ & N16,N20 \\
                2014 UM225 & o4h31 & res & $23.25\pm0.06$ & 7.21 & 44.48 & 0.10 & 18.30 & $0.79\pm0.01$ & $1.53\pm0.06$ & N16,N20 \\
                2007 TC434 & o4h39 & res & $23.21\pm0.05$ & 7.13 & 129.94 & 0.70 & 26.47 & $0.67\pm0.01$ & $1.50\pm0.06$ & N16,N20 \\
                2001 RY143 & o4h48 & cla & $23.54\pm0.08$ & 6.80 & 42.08 & 0.16 & 6.91 & $0.89\pm0.03$ & $1.89\pm0.07$ & N16,N20 \\
                2006 QP180 & o4h67PD & sca & $23.07\pm0.04$ & 9.49 & 38.08 & 0.65 & 4.96 & $0.95\pm0.08$ & $2.07\pm0.10$ & N16,N20 \\
                2014 UN228 & o4h75 & cla & $23.37\pm0.11$ & 7.46 & 45.87 & 0.17 & 24.02 & $0.59\pm0.01$ & $1.76\pm0.10$ & N16,N20 \\
                2015 RW245 & o5s06 & sca & $22.90\pm0.03$ & 8.53 & 56.48 & 0.53 & 13.30 & $0.68\pm0.02$ & $1.54\pm0.07$ & N16,N20 \\
                2004 PB112 & o5s16PD & res & $22.99\pm0.03$ & 7.39 & 107.52 & 0.67 & 15.43 & $0.82\pm0.02$ & $1.42\pm0.06$ & N16,N20 \\
                2015 RJ277 & o5s32 & cla & $23.21\pm0.04$ & 7.12 & 46.70 & 0.20 & 5.52 & $0.64\pm0.02$ & $1.18\pm0.11$ & N16,N20 \\
                2015 RG277 & o5s45 & cla & $23.15\pm0.03$ & 6.79 & 42.96 & 0.01 & 12.09 & $0.95\pm0.02$ & $1.64\pm0.06$ & N16,N20 \\
                2015 RU278 & o5s52 & res & $23.26\pm0.04$ & 6.55 & 50.85 & 0.25 & 27.20 & $0.61\pm0.03$ & $1.54\pm0.12$ & N16,N20 \\
                2015 RR245 & o5s68 & res & $21.76\pm0.01$ & 3.60 & 81.73 & 0.58 & 7.55 & $0.77\pm0.01$ & $1.19\pm0.08$ & N16,N20 \\
                2015 RU245 & o5t04 & sca & $22.99\pm0.04$ & 9.32 & 30.99 & 0.29 & 13.75 & $0.88\pm0.02$ & $1.57\pm0.07$ & N16,N20 \\
                2014 UA225 & o5t09PD & det & $22.50\pm0.02$ & 6.74 & 67.76 & 0.46 & 3.58 & $0.95\pm0.02$ & $1.46\pm0.06$ & N16,N20 \\
                2003 SP317 & o5t34PD & res & $23.48\pm0.08$ & 7.06 & 45.96 & 0.17 & 5.08 & $0.92\pm0.08$ & $1.32\pm0.13$ & N16,N20 \\
            \enddata
        \end{deluxetable}
            
        \begin{figure}
            \centering
            \includegraphics[width=\textwidth]{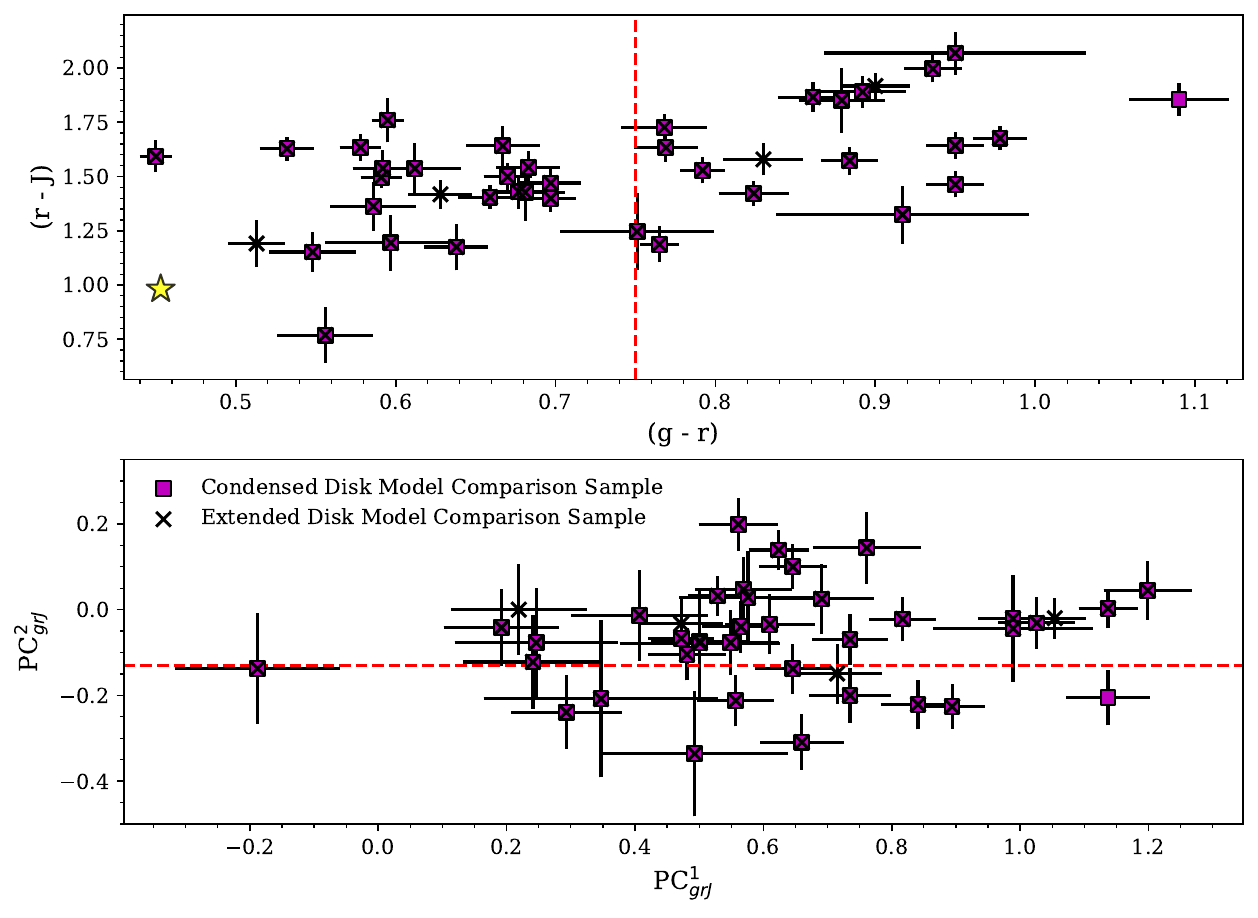}
            \caption{Col-OSSOS photometry of observed non-resonant, non-Centaur, dynamically excited objects for comparison with the two dynamical models. In the upper plot the red/neutral colour split is placed at a $(g - r)$ magnitude of 0.75. The star shows solar colours. In the lower plot the split between the BrightIR and FaintIR classes is shown by the red dashed horizontal line at $PC^{2}_{grJ}$ of -0.13.}
            \label{fig:nes16comp}
        \end{figure}
        
        \begin{table}[ht]
            \centering
            \begin{tabular}{c|cc|cc||cc|cc}
                \hline\hline
                & \multicolumn{4}{c||}{\centering \begin{tabular}[c]{@{}c@{}}Condensed Disk Model\\ Comparison Sample\end{tabular}} & \multicolumn{4}{c}{\centering \begin{tabular}[c]{@{}c@{}}Extended Disk Model\\ Comparison Sample\end{tabular}} \\
                \hline\hline
                Observing Block & Neutral    & Red    & FaintIR    & BrightIR  & Neutral   & Red    & FaintIR    & BrightIR   \\
        
                \hline
                E            & 3       & 5   & 1       & 7   &  4       & 4   & 0       & 8        \\
                L            & 5       & 1   & 2       & 4   &  5       & 1   & 2       & 4        \\
                H            & 4       & 5   & 3       & 6   &  5       & 5   & 3       & 7        \\
                O            & 6       & 0   & 0       & 6   &  6       & 1   & 0       & 7        \\
                S            & 3       & 3   & 3       & 3   &  3       & 3   & 3       & 3        \\
                T            & 0       & 3   & 3       & 0   &  0       & 3   & 3       & 0        \\
                \hline
                Total        & 20      & 17  & 11      & 26  &  23      & 17  & 11      & 29  \\
                \hline\hline
            \end{tabular}
            \caption{Summary of the numbers of the different surface types in each observing block among the the comparison sample for the condensed and extended disk models.}
            \label{tab:n_nums}
        \end{table}
    
        \subsubsection{Handling the MMRs}\label{cs16_MMRs}
        
            For removing MMR TNOs from the condensed disk model comparison sample we can use the OSSOS dynamical classifications of the observations \citep{bannister_outer_2016,bannister_ossos._2018}. Within the discovery survey for these TNOs (OSSOS) they have multiyear arcs and therefore have been accurately classified into their specific MMRs. Additionally, the final orbits of the simulated TNOs were forward integrated, and those TNOs on the major MMRs were identified \citep{nesvorny_neptunes_2016}. We therefore remove the TNOs on MMRs that are also identified by \citet{nesvorny_neptunes_2016}; the 3:2, 5:2, 4:3, 5:3, 7:4, and 2:1 MMRs. Although other higher order resonances within Col-OSSOS are identified (e.g. 9:1, 11:4 etc.) we categorise these TNOs as scattering for this work. These high order MMRs are unlikely to be contaminated with cold classicals as they did not sweep through that region during the Neptune migration phase. 
        
            When comparing with \citet{2020AJ....160...46N} we use a semimajor axis cuts for the MMRs, and we only remove the 3:2, 2:1 and 4:3 MMRs. These are the three most isolated and populated resonances, with least contamination from non-resonant TNOs. Here we use the same semimajor axis cuts as Section \ref{nes16_MMRs}. Within these limits we successfully remove all of the TNOs on each of these MMRS, and only a single additional non-resonant TNO from Col-OSSOS removed from the sample. In addition, the four TNOs on the 5:2, 5:3 and 7:4 MMRs have inclinations greater than 10$^\circ$ and are kept within the extended disk model comparison sample.
    
    \subsection{Building the Simulated Population} \label{sssec:nes16_pop}

        The dynamical models provide us with the final values for semimajor axis, eccentricity, and inclination of the simulated TNOs. However, to effectively utilise the OSSOS survey simulator and make a meaningful comparison with Col-OSSOS, we need more comprehensive information, including the complete orbits, $H_r$ magnitudes, and positions of the simulated TNOs. Therefore, it is necessary to generate distributions for the orbital angles, such as longitude of ascending node, argument of pericenter, and mean anomaly, as well as values for the $H_r$ magnitudes. We need a large number of simulated TNO detections in each observing block, however we must avoid modifying the final orbits of the simulated TNOs, namely their semimajor axis, eccentricity, and inclination, as these values depend on the initial positions of the TNOs. Therefore, we address this challenge by artificially increasing the number of simulated TNOs. This involves replicating the final orbits obtained from the dynamical model and subsequently randomly assigning orbital angles and absolute magnitudes from the relevant distributions to these duplicated orbits.
        
        The $H_r$ magnitude distribution is a broken exponential with a sharp transition \citep{fraser_absolute_2014,shankman_ossos_2016,lawler_ossos_h._2018}. The bright end of this distribution has a steeper slope than the faint end, with the break at a $H_r$ magnitude of 7.7 \citep{shankman_ossos_2016,lawler_ossos_h._2018}. The $H_r$ distribution follows Equation \ref{eqn4:H_pre} at the bright end, where $H_r < 7.7$. Here $N(\leq H)$ is the cumulative number of TNOs at magnitude $H$, $H_0 = 3.6$ is a normalisation constant with value equal to the brightest $H_r$ magnitude for OSSOS detections, and $\alpha_1 = 0.9$ \citep{fraser_absolute_2014}. The faint end of the $H_r$ magnitude distribution follows Equation \ref{eqn4:H_post}. Here $H_B$ is the magnitude at the break in the distribution \citep[$H_B = 7.7;$][]{lawler_ossos_h._2018}, the contrast value $c = 0.85$, and $\alpha_2 = 0.4$ \citep[][]{fraser_absolute_2014}.
    
        \begin{equation}
            \label{eqn4:H_pre}
            N(\leq H) = 10^{\alpha_1 (H_r - H_0)}
        \end{equation}
    
        \begin{equation}
            \label{eqn4:H_post}
        \begin{split}
            N(\leq H) = c 10^{\alpha_1 (H_B - H_0)} + &B 10^{\alpha_2 (H - H_B)} -B \\
            B = c \frac{\alpha_1}{\alpha_2} &10^{\alpha_1(H_B - H_0)}
        \end{split}
        \end{equation}

        The $H_r$ magnitudes that we generate range between $3 < H < 11$. Brighter than  $H_r$ magnitude of 3 the distribution changes due to nearing the sizes dwarf planets \citep[which have different compositions and brightnesses than smaller TNOs,][]{2008ssbn.book..335B}. Additionally, within OSSOS no TNOs were observed brighter than $H_r \sim 3$. For the faint end cut-off, within Col-OSSOS there were no TNOs with $H_r$ dimmer than 11 and few within OSSOS. We therefore limited our simulated TNOs to $H_r$ magnitudes less than 11 so as to reduce the needed computing time. 

        For the orbital angles of the simulated TNOs (longitude of ascending node, argument of pericenter, and mean anomaly) we draw randomly from a uniform distribution between $0^\circ$ and $360^\circ$. These angles have been uniformly randomised by planetary effects over the past $\sim$4 billion years since the end of Neptune's migrations \citep{brasser_embedded_2006}. Along with the absolute $H_r$ magnitudes and the orbits from the condensed disk model \citep{nesvorny_neptunes_2016} or the extended disk model \citep{2020AJ....160...46N} we can fully simulate the modelled Kuiper belt population.
            
    \subsection{Running the Colour Simulations}\label{ssec4:n_csims}

        After creating the various orbital distributions we then created simulated populations to input into the OSSOS survey simulator \citep{lawler_ossos:_2018}. In order to create a large number of simulated `detections' we created large numbers of the simulated TNOs; using the orbits from the dynamical models after the dynamical cuts in Section \ref{ssec4:dyn_cuts} and drawing the orbital angles and absolute magnitudes from the distributions in Section \ref{ssec4:col_n}. This is essentially a simulation of OSSOS, and so by limiting the brightnesses to $m_r <23.6$ and only including detections in the Col-OSSOS observing blocks (E, L, H, O, S and T) we simulate the selection of the Col-OSSOS targets. Finally, we assign red/neutral colours or BrightIR/FaintIR surfaces to the simulated TNOs based on their pre-Neptune migration positions. For each transition positions we ran the simulations until we had 1000 Col-OSSOS detections in each Col-OSSOS observing block. This exact number of 1000 detections is relatively arbitrary, we simply wanted a large number of detections per block.
        
        To determine the most plausible scenario consistent with the observed surface properties of the Kuiper belt from Col-OSSOS, we employ binomial probabilities. These probabilities are calculated for each individual observing block and are denoted as $P(r)$, they quantify the likelihood of the observed data aligning with the predictions generated by the colour simulations. Equation \ref{eqn:poisson} describes the probability.
        
        \begin{equation}
            \label{eqn:poisson}
            P(r) = \frac{n!}{(n-r)!r!}p^r(1-p)^{(n-r)}
        \end{equation}

        In this equation, the variable $n$ represents the total number of simulated TNOs, while $r$ corresponds to the number of TNOs with a specific surface characteristic (either red or FaintIR) within a particular observing block. The term $p = N_r/N$ signifies the ratio of TNOs with the specific surface characteristic (either red or FaintIR) within the Col-OSSOS comparison sample observing block, where $N_r$ denotes the count of TNOs with the specified surface characteristic. \revision{Because each Col-OSSOS observing block samples a distinct and non-overlapping region of the sky, the colour distributions derived from different blocks represent independent measurements that can be directly combined in the statistical analysis. Therefore, to} determine the overall probability across all Col-OSSOS observing blocks, we simply multiply the individual binomial probabilities of the six blocks together.

        The combined probability, denoted as $P(r)_{combined}$, represents the combination of the individual binomial probabilities from each observing block. The intention behind calculating the individual probabilities for each observing block and subsequently combining them was to ensure that no single block exerted a dominant influence on the overall signal due to their distinct characteristics. Consequently, the resulting $P(r)_{combined}$ values tend to be relatively small, as we are multiplying probabilities that are individually less than 1. However, since our focus is on investigating the peak, the specific magnitudes of these low probability values are inconsequential. Alternatively, we could have computed a total binomial probability by considering the total number of surfaces in the Col-OSSOS comparison sample, instead of combining the individual block probabilities. This approach would have yielded higher probabilities; however, it would have sacrificed valuable information by neglecting the constraints imposed by the individual block pointings in our results.

    \subsection{Calculated Transition Positions} \label{ssec4:nes_results}

\begin{figure}[ht]
            \centering
            \includegraphics[width=\textwidth]{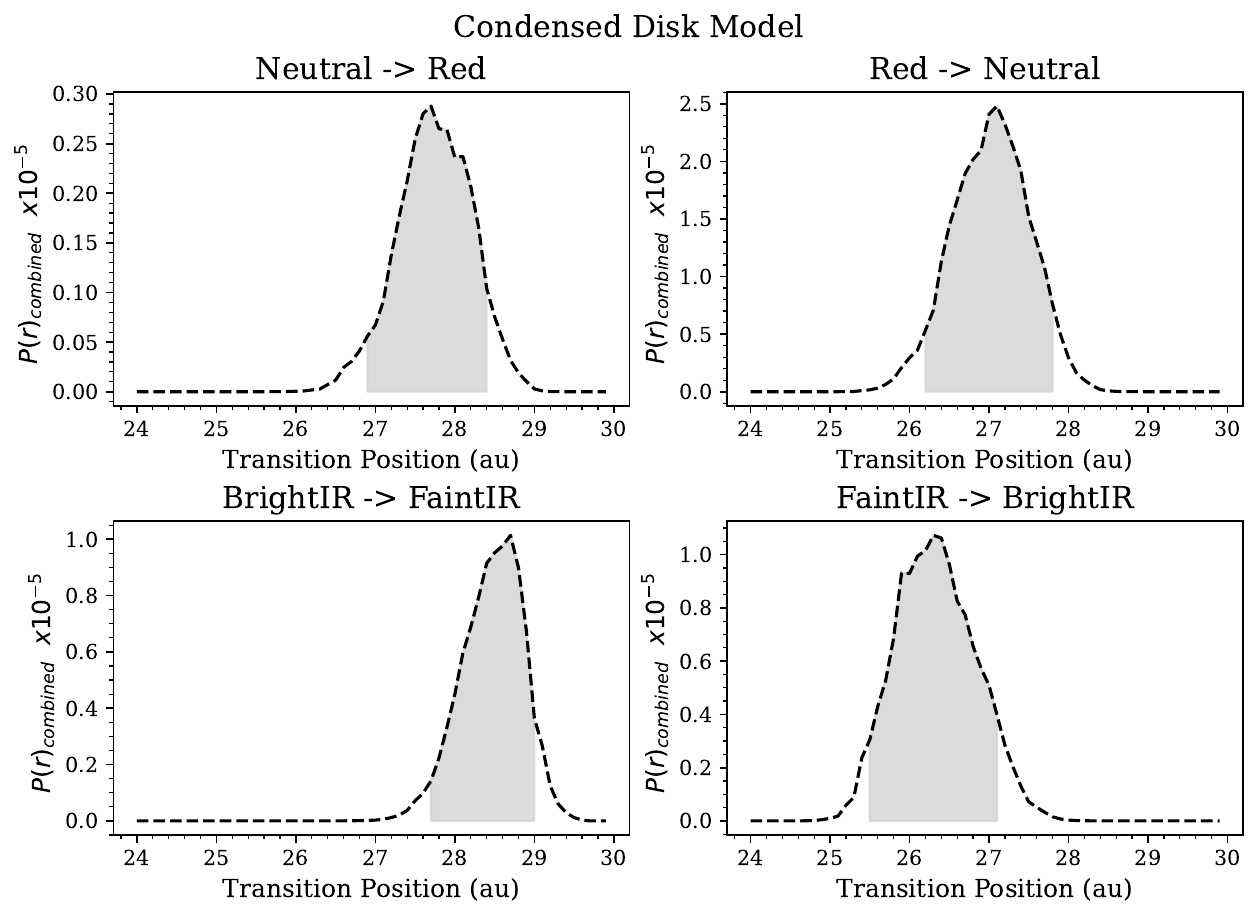}
            \caption{How the binomial probability changes with the position of the synthetic ice line position in the primordial planetesimal disk. The initial disk ranges from 24 to 30.36 au. The upper plots are for a red/neutral surface classification, while the lower plots are for a BrghtIR/FaintIR classification. The subplot titles show the specific primordial disk layouts. The grey shaded region shows the 95\% area under the curve.}
            \label{fig:nes16_rbbr}
        \end{figure}

        Table \ref{tab:conc_results} shows the most likely transition position for each of the disk layouts we investigated. Figure \ref{fig:nes16_rbbr} plots the $P(r)_{combined}$ obtained from the simulations of the red/neutral condensed disk model, plotted against the red/neutral transition position in the primordial disk. For the inner neutral/outer red initial disk, we determine that the transition position is $27.7^{+0.7}_{-0.8}$ au, with the peak of the distribution serving as the best-fit value. To establish the upper and lower limits, we calculate the transition position values corresponding to the 95\% limit of the area under the curve. To precisely define the boundaries of the region encompassing the peak of the distribution, we conduct additional simulations with a step size of 0.1 au. This process allows us to accurately narrow down the edges of the region surrounding the peak. Similarly, for the inner red/outer neutral disk, we find that the colour transition position is $27.1^{+0.8}_{-0.8}$ au. Additionally, we have assigned BrightIR and FaintIR surfaces \citep[from][]{2023PSJ.....4...80F} to the initial disk, in order to investigate where in the primordial solar system these surfaces originated. Similar to the red/neutral surfaces we take the peak in the distribution as the most likely transition position, and 95\% area limits to find the 95\% confidence limits. This results in a transition position of $28.7^{+0.3}_{-1.0}$ au for the inner BrightIR / outer FaintIR primordial disk, and $26.3^{+0.8}_{-0.8}$ au for the inner FaintIR / outer BrightIR disk.

        \begin{table}[ht]
            \centering
            \begin{tabular}{c|c|c}
                \hline\hline
                                               & \multicolumn{2}{c}{Transition Position}     \\
                \hline
                Disk Layout                    & Condensed Disk Model & Extended Disk Model  \\
                \hline\hline
                Inner Neutral / Outer Red      & $27.7^{+0.7}_{-0.8}$ au & $30.0^{+1.1}_{-1.2}$ au \\
                Inner Red / Outer Neutral      & $27.1^{+0.8}_{-0.8}$ au & $28.6^{+1.1}_{-1.2}$ au \\
                Inner BrightIR / Outer FaintIR & $28.7^{+0.3}_{-1.0}$ au & $31.5^{+1.1}_{-1.2}$ au \\
                Inner FaintIR / Outer BrightIR & $26.3^{+0.8}_{-0.8}$ au & $26.9^{+1.4}_{-1.1}$ au \\
                \hline\hline
            \end{tabular}
            \caption{Summary of the transition positions for the different scenarios investigated in Section \ref{sec:colourtransitions}.}
            \label{tab:conc_results}
        \end{table}
    
\begin{figure}[ht]
            \centering
            \includegraphics[width=\textwidth]{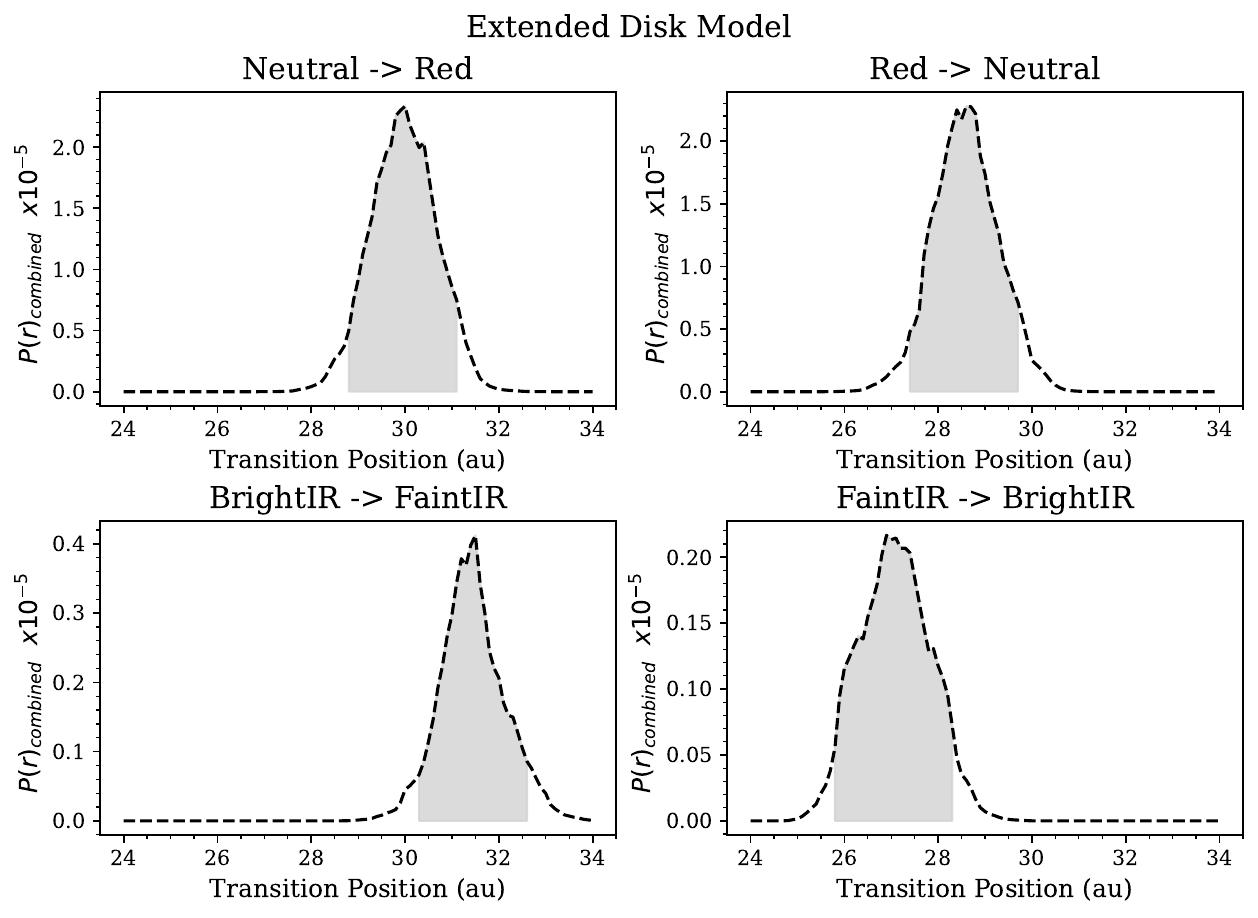}
            \caption{How the binomial probability changes with the position of the synthetic ice line position in the primordial planetesimal disk. The initial disk ranges from 24 to 50 au, however we only show 24 au to 34 au here so as to zoom in on the transition area. The upper plots are for a red/neutral surface classification, while the lower plots are for a BrghtIR/FaintIR classification. The subplot titles show the specific primordial disk layouts. The grey shaded region shows the 95\% area under the curve.}
            \label{fig:nes20_rbbr_tot}
        \end{figure}

        In Figure \ref{fig:nes20_rbbr_tot} we show the red/neutral binomial probabilities of the \citet{2020AJ....160...46N} extended disk migration model, as a function of the colour transition position in the primordial disk. We find that for the inner neutral / outer red initial disk the transition position is $30.0^{+1.1}_{-1.2}$ au, using the peak of the distribution as the best fit value 95\% area under the curve limits to find the uncertainty. Similarly, we find that for the inner red / outer neutral disk the colour transition position is $28.6^{+1.1}_{-1.2}$ au. In the region of the distribution peak we ran additional simulations with a 0.1 au step size in order to accurately narrow down the edges of the region. Additionally, we investigate the BrightIR / FaintIR formation positions. Figure \ref{fig:nes20_rbbr_tot} shows the results of these simulations. Similar to the red/neutral surfaces we take the peak in the distribution as the most likely transition position, and use  the 95\% area under the curve to find the limits. This results in a transition position of $31.5^{+1.1}_{-1.2}$ au for the inner BrightIR / outer FaintIR primordial disk, and $26.9^{+1.4}_{-1.1}$ au for the inner FaintIR / outer BrightIR disk.

        In addition to finding the most probable transition position, we have observational constraints that we can discuss in the context of the transition position. First we have the `blue binaries', these are wide binaries with neutral coloured and BrightIR categorised surfaces which reside on cold classical orbits. Simulations suggest that they could not survive a long migration without losing their binarity. Therefore, \citet{fraser_all_2017} suggest that these binaries would have formed at $\sim$38 au. \citet{2022AJ....163..137N} further investigate this however, and find that all but the widest of these `blue binaries' can survive a push out from $<$30 au (although they will have higher survival rates if forming further out). This implies that all of our inner red / outer neutral and inner FaintIR / outer BrightIR simulations are consistent with the formation of the `blue binaries'. Additionally, the inner neutral / outer red extended disk model simulations are also consistent with the theorised formation positions of the `blue binaries'. 

        The colours exhibited by Neptune Trojans provide an additional constraint on the formation locations of different TNOs. In a recent study, \citet{2023MNRAS.521L..29B} revealed that 27 out of 32 Neptune Trojans displayed neutral-coloured surfaces. To account for the presence of a significant number of red-surfaced TNOs with orbits interior to Neptune, \citet{2023MNRAS.521L..29B} proposed that the transition position between an inner neutral / outer red primordial disk should be situated within the range of 30 au to 35 au. This finding aligns with the outcomes of all our simulations that feature an inner neutral / outer red configuration. However, since the colours of the Neptune Trojans have not been measured in the $J-$band, we are unable to assess the consistency of the BrightIR/FaintIR simulations with the observed colours of the Neptune Trojans. Additionally \citet{2023MNRAS.521L..29B} do not comment on an inner red / outer neutral disk configuration so we cannot compare these simulations with their work. 

        Lastly, it is known that neutral-coloured TNOs, as well as BrightIR TNOs, exhibit higher inclinations and eccentricities compared to the FaintIR or red TNOs \citep{2019AJ....157...94M,2021AJ....162...19A,2023PSJ.....4..160M}. Furthermore, \citet{2020AJ....160...46N} demonstrated that in the extended disk model, TNOs situated on interior (or more sunward) orbits acquire greater inclinations during Neptune's migration than those formed further out. This suggests that the configurations of inner neutral/outer red and inner BrightIR/outer FaintIR primordial disks are more likely to yield orbits consistent with the observed characteristics of each colour class. To investigate this in the context of our study, we apply Mann-Whitney U tests \citep{MannWhitney} to the post-migration inclinations and eccentricities, combining the results using Fisher's method \citep{fisher1970statistical}. Our null hypothesis states that planetesimals originating closer to the Sun in the pre-migration disk will, on average, attain higher inclinations and eccentricities than those that started farther out. If we can reject this null hypothesis for a given disk layout, we can rule out that layout as a potential explanation for the inclination/eccentricity colour trends observed in the Kuiper Belt.

        We focus on the most probable surface transition for each disk layout, as listed in Table \ref{tab:conc_results}. For each layout, we generate $10^5$ subsamples of test particles of the same size as the observed populations (37 for the condensed disk model and 40 for the extended disk model). Each subsample undergoes two separate Mann-Whitney U tests: one testing whether planetesimals that started closer to the Sun tend to reach higher inclinations and another testing whether they attain higher eccentricities. We then combine the p-values from these two tests using Fisher's combined probability method. Table \ref{tab:2dks} presents the percentage of the $10^5$ subsamples that resulted in a combined Fisher p-value of less than 0.05, indicating the proportion of cases in which we reject the null hypothesis. If this percentage is below 5\%, we conclude that the null hypothesis can be rejected. Our analysis shows that we can reject the null hypothesis for all disk layouts except the inner neutral / outer red configurations in both the condensed and extended models, as well as the inner BrightIR / outer FaintIR configuration in the extended disk model.
        
        \begin{table}[ht]
            \centering
            \begin{tabular}{c|c|c}
                \hline\hline
                                               & \multicolumn{2}{c}{Fisher's Combined Probability}     \\
                \hline
                Disk Layout                    & Condensed Disk Model & Extended Disk Model  \\
                \hline\hline
                Inner Neutral / Outer Red      & 6\% & 24\% \\
                Inner Red / Outer Neutral      & 2\% & 1\% \\
                Inner BrightIR / Outer FaintIR & 4\% & 25\% \\
                Inner FaintIR / Outer BrightIR & 4\% & 1\% \\
                \hline\hline
            \end{tabular}
            \caption{Summary of the Fisher's combined probabilities for each initial disk scenario, combining Mann-Whitney U tests with the null hypothesis that the inner starting surface composition produces higher inclinations and eccentricities than the outer surface composition. If the Fisher's combined probability is $<5\%$ the null hypothesis can be rejected.}
            
            \label{tab:2dks}
        \end{table}

        \revision{Finally, in order to assess how well the models reproduce the observed surface colour distribution, we compared the fraction of red or FaintIR objects within each dynamical class (simplified to classical as defined in Section \ref{ssec4:dyn_cuts} and all other objects simply categorised as scattered) across the six observing blocks. We quantified the agreement using the Pearson correlation coefficient between model-predicted and observed red or FaintIR fractions. For the disks with inner neutral / outer red and inner BrightIR / outer FaintIR initial disk cases, we find a moderate positive correlation (Pearson coefficient $\sim$0.5) for both dynamical classes, indicating that the models reasonably capture the observed spatial colour trends. In contrast, for the inner red / outer neutral and inner FaintIR / outer BrightIR initial disk cases, classical objects continue to show a similar moderate positive correlation, whereas scattered objects exhibit little to no correlation. This suggests that the inner red and inner FaintIR disk models are less successful at reproducing the colours of the more distant/scattered TNOs observed by Col-OSSOS.}
        
\section{Estimating the True Colour Fraction in the Kuiper Belt} \label{sec4:L7sims}

    After investigating the origins of various surface types in the Kuiper belt through dynamical models, we proceed to estimate the proportions of different surface types using model of the true population of the Kuiper belt today. This model is based on the Canada-France Ecliptic Plane Survey (CFEPS) \citep{2006Icar..185..508J,2011AJ....142..131P,2017AJ....153..236P}. To compare with the Col-OSSOS observations, we apply colour ratios (red and neutral) or surface ratios (BrightIR and FaintIR) to the intrinsic Kuiper belt model and incorporate biases using the OSSOS survey simulator. This enables us to establish meaningful comparisons between the model and the Col-OSSOS dataset. In this study, we utilise the orbits generated by the CFEPS L7 model \citep{2011AJ....142..131P} and assign colours to the TNOs based on a chosen ratio. Specifically, we focus on analysing the dynamically hot simulated TNOs. The cold classicals do not have the surface optical colour bimodality of the dynamically excited population and so do not add anything to the analysis. Additionally, the H magnitude distribution for the CFEPS L7 is not accurate, as described in Section \ref{ssec4:L7}. By employing the OSSOS survey simulator, we simulate the detection of coloured TNOs while applying the 23.6 $r$-band magnitude threshold used in Col-OSSOS. This allows us to compare the colours observed by Col-OSSOS with those generated by the intrinsic colour simulations. To determine the best match, we calculate binomial probabilities that assess the likelihood of the observed colours aligning with the simulated colours.
    
    \subsection{CFEPS L7 Model}\label{ssec4:L7}
    
        The CFEPS L7 model provides debiased measurements of the intrinsic populations within the Kuiper belt, encompassing the classical, scattering, detached, and resonant populations. Initially introduced in \citet{2011AJ....142..131P}, the model resonant populations were further refined in \citet{2012AJ....144...23G}. The intrinsic model was designed to align the bulk of its orbital properties with those found in the CFEPS orbit catalogue. It should be noted that, due to limitations in survey detectability, the model's accuracy might be compromised in areas with low detectability. Nevertheless, it currently stands as the only debiased model of the full Kuiper belt and has been widely employed in various previous studies. Utilising the CFEPS L7 model, we have access to the complete orbits, orbital angles, and $H_g$ magnitudes of the simulated TNOs.
    
        Although both the hot and cold classical TNOs are included in this model, the absolute magnitude distribution for their cold classical TNOs is different than that of the hot population. It has since been found that this distinction is inaccurate \citep{2023ApJ...947L...4P}, and therefore the $H$ magnitude for the cold classicals in this model is incorrect. However, there is no impact on our analysis as we simply compare the dynamically hot population as in Section \ref{ssec4:dyn_cuts} and therefore we have removed the cold classicals from this model. We remove the simulated TNOs in the classical region (semimajor axis range 37.37 au $< a <$ 47.7 au, eccentricities below 0.24) with inclinations below $5^\circ$.

    \subsection{Col-OSSOS Comparison Sample}\label{ssec:L7_comp}
    
        For accurate comparisons between the model and the observations via the survey simulator we must again create a Col-OSSOS comparison sample. This is summarised below and follows the same method as Section \ref{ssec4:col_n}, though including the MMRs this time. We only include TNOs with $m_r < 23.6$, as this is the same discovery magnitude cut that we apply to the L7 model as described in Section \ref{sec:colourtransitions}. For all but 2013 JE64 of the repeat observations we use the initial optical colour measurement to categorise their surfaces. For 2013 JE64 we use the follow-up optical photometry to categorise the TNO as neutral colours, due to the initial observation being impacted by a bright background star and therefore unreliable. We also remove the Haumea collisional family member from the sample (2013 UQ15) as this TNO's surface colour is the result of a collision and is therefore not primordial. We remove the cold classical TNOs as with the L7 model. This involves taking TNOs within the classical belt of semimajor axis range 37.37 au $< a <$ 47.7 au and eccentricities below 0.24, and removing those with inclinations below $5^\circ$. We summarise the Col-OSSOS comparison sample for the L7 model in Table \ref{tab:L7_nums}.
        
        \begin{table}[ht]
        \centering
            \begin{tabular}{c|cc|cc}
                \hline\hline
                Observing Block & Neutral    & Red    & FaintIR    & BrightIR  \\
                \hline
                E               &  8       & 5   & 1       & 12       \\
                L               &  5       & 3   & 3       & 5        \\
                H               &  14      & 6   & 6       & 14       \\
                O               &  7       & 5   & 3       & 9        \\
                S               &  6       & 3   & 3       & 6        \\
                T               &  0       & 4   & 4       & 0        \\
                \hline
                Total           &  40      & 26  & 20      & 46   \\
                \hline\hline
            \end{tabular}
            \caption{Summary of the numbers of different surfaces among the the Col-OSSOS comparison sample for the L7 model.}
            \label{tab:L7_nums}
        \end{table}
    
        \subsection{CFEPS L7 Synthetic Colour Simulations}\label{ssec4:L7sims}
        
            After the dynamical cuts in Section \ref{ssec4:L7} we can input the simulated TNOs into the OSSOS survey simulator. This biases the intrinsic Kuiper belt populations with its colour ratio to the TNOs OSSOS would have detected. We assign red and neutral colours to the TNOs based on the colour ratio that we are investigating. For the $(g-r)$ values we use the mean $(g-r)$ for the red and neutral groups in the comparison sample of Col-OSSOS TNOs. Therefore we use $(g-r) = 0.896$ for the red TNOs, and $(g-r) = 0.628$ for the neutral coloured TNOs.
    
            Similarly to our investigation in Section \ref{ssec4:n_csims}, we run the simulations until we have a total of 1000 `detections' with $m_r<23.6$ in each observing block. Although the L7 model uses $H_g$ absolute magnitudes, we set the survey simulator detection filter to $r-$band and provide the surface colours so that the discovery magnitude ($m_r$) can be adjusted to $r-$band. Following the completion of our 1000 `detections' per block, we proceed to calculate the frequency of simulated TNOs exhibiting red or FaintIR surfaces. Utilising Equation \ref{eqn:poisson}, we determine the binomial probability associated with this intrinsic colour ratio, which signifies the likelihood of producing the observed colours observed by Col-OSSOS.

    \subsection{Intrinsic Surface Fractions in the Kuiper Belt} \label{sec4:L7results}
    
\begin{figure}[ht]
            \centering
            \includegraphics[width=\textwidth]{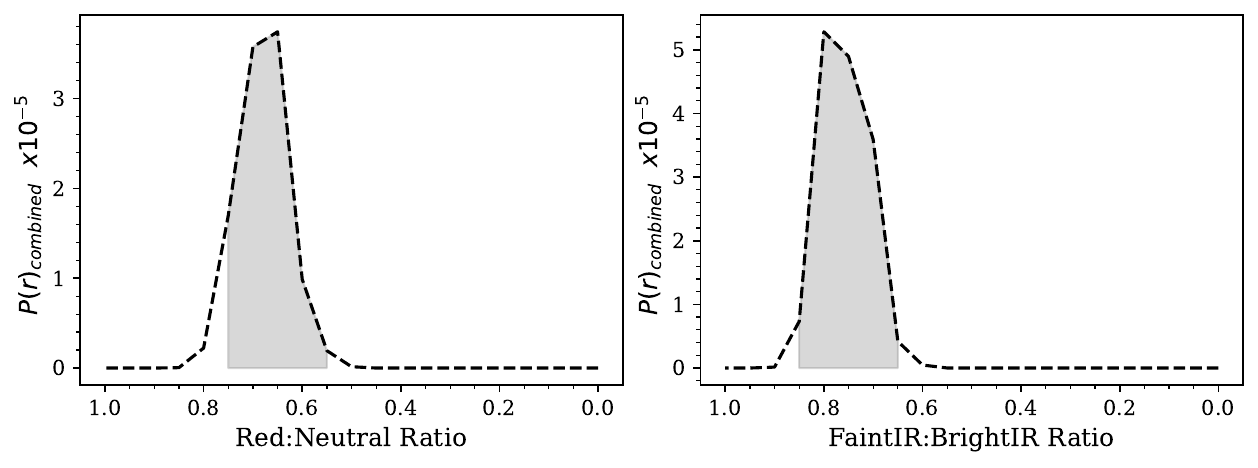}
            \caption{How the binomial probability changes with the surface ratio in the Kuiper belt. The left panel shows the results for the red/neutral ratios, and the right panel shows the results for the BrightIR/FaintIR ratios. The grey shaded region shows the 95\% area under the curve.}
            \label{fig:L7}
        \end{figure}

        In Figure \ref{fig:L7}, we present the variation of binomial probabilities in relation to the ratio of different surface types in the Kuiper belt. The peak of the distribution represents the most probable intrinsic colour ratio that could account for the observed colours in the Col-OSSOS data. Similar to the methodology outlined in Section \ref{ssec4:nes_results}, we determine the limits on this result by calculating the bounds that encompass 95\% of the area under the curve. Our analysis reveals that the red-to-neutral surface ratio ranges from 0.33:1 to 0.82:1, with the peak value occurring at a ratio of 0.54:1 (35:65 in Figure \ref{fig:L7}). Previous studies \citep{schwamb_col-ossos:_2019,2017AJ....153..145W} have reported a slightly lower red-to-neutral ratio of 0.1-0.3:1. However, it is important to note that these prior investigations examined the colour ratio within a size limit rather than a magnitude limit. Considering that neutral-coloured TNOs generally have lower albedos than those with redder colours \citep{lacerda_albedo-color_2014}, it is expected that the red-to-neutral ratio will be higher when working within an magnitude regime. We find that the FaintIR:BrightIR ratio varies between 0.18:1 to 0.54:1, with the peak at 0.25:1, as can be seen in Figure \ref{fig:L7}.

\section{Conclusions} \label{sec4:conclusions}

    We have investigated the surface colours and ratios of TNOs in the Kuiper belt, exploring the origins and distributions of different surface types. By utilising dynamical models \citep{nesvorny_neptunes_2016,2020AJ....160...46N} we investigated the origin positions of different surface types in the early solar system. With the intrinsic Kuiper belt model based on CFEPS L7 model, we were able to study the surface colour ratios and compare them with the observations from Col-OSSOS. 
    
    In our investigation of different primordial disk compositions we found a range of scenarios that are consistent with the colours observed by Col-OSSOS today, with the final transition positions shown in Table \ref{tab:conc_results}. Additionally, if we take into account further observational factors such as the `blue binaries' or Neptune Trojans we can comment about the likelihood of specific disk layouts. For instance, in the condensed disk model, the inner red / outer neutral disk configuration is unlikely to be consistent with the formation of the observed `blue binaries'. Additionally, both inner red / outer neutral disks align well with the proposed colour constraints for Neptune Trojans as suggested by \citet{2023MNRAS.521L..29B}. Finally, neutral coloured and BrightIR TNOs are found at higher inclinations than their counterparts \citep{2019AJ....157...94M,2021AJ....162...19A,2023PSJ.....4..160M}. To further investigate the correlation between inclination and surface colour, we combined two Mann-Whitney U tests using Fisher's method. Based on this we could rule out all but the inner neutral / outer red condensed and extended disk models, as well as the inner BrightIR / outer FaintIR extended disk model. Previously, \citet{2020AJ....160...46N} showed that simulated TNOs that started on interior orbits gained higher inclinations than those that formed further out. Additionally, \citet{2023PSJ.....4...80F} propose that the TNO surfaces are made up of solely BrightIR and FaintIR, including the cold classical TNOs that are predominantly FaintIR and formed further out in the early solar system. Together these suggest that the inner neutral / outer red and the inner BrightIR / outer FaintIR for particularly the extended disk model are the most likely configurations of the primordial disk.
    
    Overall, our findings demonstrate that there exist multiple viable scenarios for the composition of primordial disks that can explain the observed colours of TNOs in the Col-OSSOS survey. While certain disk layouts may be less likely when considering additional observational constraints, a range of possibilities remains open. Considering the potential `most likely transitions' of a transition $30.0^{+1.1}_{-1.2}$ au for the inner neutral / outer red, and a transition of $31.5^{+1.1}_{-1.2}$ au for inner BrightIR / outer FaintIR (both for the extended disk model) we can speculate on their interpretation. This might suggest that the cold classical TNOs (with their red/FaintIR surfaces) formed simply as an extension of the massive planetesimal disk from which the dynamically hot population originated. Recent JWST observations of TNO surface spectra \citep[e.g.,][]{2024NatAs.tmp..305D} suggest that cold classical TNOs share a carbon-ice-rich surface composition with the redder dynamically excited TNOs. This similarity may be due to their more distant formation regions, which allowed them to retain carbon ices over those TNOs that formed further sunward. Over time, irradiation of these ices likely led to the formation of complex hydrocarbons, contributing to their distinctly red surface colours. Additionally, the surface transition occurring at approximately the same position as the drop in surface density might suggest some difference in collisional evolution causing the different surface types. 
    
    Finally, we examined the colour ratios within the Kuiper belt by applying a colour ratio to a model of the intrinsic orbit distribution (CFEPS L7 model) and biasing it with the OSSOS survey simulator. Our results showed that the most likely red:neutral colour ratio is 0.54:1, though ranging from 0.33:1 to 0.82:1. These findings suggest a higher red:neutral ratio than previously reported studies, which focused on colour ratios within specific size limits rather than absolute magnitude regimes. We also investigated the FaintIR:BrightIR colour ratio, and found the most likely ratio to be 0.25:1, ranging from 0.18:1 to 0.54:1, though this ratio will tend to favour a higher fraction of  neutral/BrightIR when albedos are considered.
    
    It is worth noting that our study was based on the best available models and data at the time. While we took into account potential limitations, such as the accuracy of the intrinsic Kuiper belt model and the distinction between hot and cold classical TNOs, further advancements in observations and modelling may refine our understanding of the Kuiper belt's surface colours and their origins. Additionally, the recent Dark Energy Survey \citep{2025arXiv250101551B} use a larger sample of TNOs with less precise photometric colours to postulate on the origin of different surfaces within the Kuiper belt. They confirm that it is very likely that FaintIR and BrightIR populations originated in different areas in the solar system, although do not investigate their precise formation location in the context of dynamical models of Kuiper belt formation. In the future we will have further means to examine the colour ratios and primordial colour distributions in significantly further detail; over the next decade or so the LSST will increase the number of known TNOs by a factor of 10. The optical colours measured by this survey will provide the means to further this work on a much larger scale. While LSST will improve sample sizes, our analysis demonstrates that meaningful progress can be made with existing data. The developed methodology presented in this work is ready to be applied to upcoming dataset. Therefore, it is relevant to present it to the community who will be interested to apply it to upcoming data.
    
\begin{acknowledgments}
    LEB acknowledges funding from the Science Technology Facilities Council (STFC) Grant Code ST/T506369/1 and the Can-Rubin Fellowship at the University of Victoria. MTB appreciates support during OSSOS from UK STFC grant ST/L000709/1, the National Research Council of Canada, and the National Science and Engineering Research Council of Canada. KV acknowledges support from NASA (grants 80NSSC21K0376 and 80NSSC23K1169). MES acknowledges support from STFC grant ST/V000691/1. REP acknowledges NASA Emerging Worlds grant 80NSSC21K0376 and NASA Solar System Observations grant 80NSSC21K0289. This work was supported by the Programme National de Plantologie (PNP) of CNRS-INSU co-funded by CNES. The authors acknowledge the sacred nature of Maunakea and appreciate the opportunity to obtain observations from the mountain.  The observations were obtained as part of observations from the programs (GN-2014B-LP-1, GN-2015A-LP-1, GN-2015B-LP-1, GN-2016A-LP-1, GN-2017A-LP-1, GN-2018A-Q-118, GN-2018A-DD-104, GN-2020B-Q-127, and GN-2020B-Q-229) at Gemini Observatory. The international Gemini Observatory is a program of NSF’s OIR Lab, and is managed by the Association of Universities for Research in Astronomy (AURA) under a cooperative agreement with the National Science Foundation. On behalf of the Gemini Observatory partnership: the National Science Foundation (United States), National Research Council (Canada), Agencia Nacional de Investigación y Desarrollo (Chile), Ministerio de Ciencia, Tecnología e Innovación (Argentina), Ministério da Ciência, Tecnologia, Inovações e Comunicações (Brazil), and Korea Astronomy and Space Science Institute (Republic of Korea). The GMOS-N observations were acquired through the Gemini Observatory Archive at NSF’s NOIRLab and processed using DRAGONS (Data Reduction for Astronomy from Gemini Observatory North and South). We are grateful for use of the computing resources from the Northern Ireland High Performance Computing (NI-HPC) service funded by EPSRC (EP/T022175). Data Access: data supporting this study are included within the article and the Gemini Observatory Archive (\href{https://archive.gemini.edu}{https://archive.gemini.edu}) This research used the Canadian Advanced Network For Astronomy Research (CANFAR) operated in partnership by the Canadian Astronomy Data Centre and The Digital Research Alliance of Canada with support from the National Research Council of Canada the Canadian Space Agency, CANARIE and the Canadian Foundation for Innovation.
\end{acknowledgments}

\vspace{5mm}
\facilities{Gemini North (GMOS, NIRI)}

\software{numpy \citep{harris2020}, matplotlib \citep{hunter2007}, TRailed Imaging in Python, \citep{2016AJ....151..158F}, astropy \citep{2013A&A...558A..33A,2018AJ....156..123A}, scipy \citep{2020SciPy-NMeth},
          SExtractor \citep{1996A&AS..117..393B}, OSSOS Survey Simulator \citep{lawler_ossos:_2018,2018ascl.soft05014P}}

\bibliography{sample631}{}
\bibliographystyle{aasjournal}

\end{document}